\documentclass[11pt,a4paper]{article}

 \usepackage{cite}
\usepackage[english]{babel}
\usepackage{amsmath}
\usepackage{amsfonts}
\usepackage{amssymb}
\usepackage{lmodern}
\usepackage[left=2.5cm,right=2.5cm,top=2.5cm,bottom=2.5cm]{geometry}
\usepackage{tensor}
\usepackage{braket}
\usepackage{slashed}
\usepackage{graphicx}
\usepackage{xcolor}
\usepackage{mathrsfs} 

\usepackage{caption}
\usepackage{hyperref} %
\hypersetup{
bookmarksnumbered=true,
    colorlinks=true, %
    linkcolor=black, %
    urlcolor=black, %
    citecolor = black,
    linktoc=all %
}

\allowdisplaybreaks  %

\newcommand{\DD}[1]{\mathcal{D}#1\ }
\newcommand{\dd}[1]{d#1\ }

\newcommand{\gr}{ { \text{grav} } }
\newcommand{\Riem}{ { \text{Riem} } }
\newcommand{\Ric}{ { \text{Ric} } }

\newcommand{\cd}{ { \mathcal{\nabla} } }

\newcommand{\EE}{ \mathbb{E}_4 }
\newcommand{\WW}{ \mathrm{Weyl}^2 }
\newcommand{\An}{ \mathcal{A} }

\newcommand{\dm}[1]{ {d \hspace{-0.5em} \overline{\phantom {x}} } #1 \, }
\newcommand{\bb}[1]{ \braket {\!  \braket{  #1  } \! } }

\title{Conformal Anomalies and Renormalized stress tensor correlators}
\author{Tommaso Bertolini and Lorenzo Casarin}
\date{\today}

\begin{document} 
\begin{center}
\vspace*{1cm}

{\Large\bf {
Conformal `anomalies'
\\
and renormalized stress-tensor correlators
\\[0.4em]
for non-conformal theories 
}
}

\vspace{1cm}

{Tommaso Bertolini\textsuperscript{1} 
and
Lorenzo Casarin\textsuperscript{2,3} 
}

\vspace{0.5cm}

{

\textsuperscript{\rm 1}Department of Physics \\ Stockholm University \\ AlbaNova University Center, 106 91, Stockholm, Sweden \\
\vspace{0.5cm}
\textsuperscript{\rm 2}Institut f\"ur Theoretische Physik  \\ 
Leibniz Universit\"at Hannover \\
Appelstra\ss{}e 2, 30167 Hannover, Germany \\
\vspace{0.5cm}
\textsuperscript{\rm 3}Max-Planck-Institut f\"u{}r Gravitationsphysik (Albert-Einstein-Institut)  \\
Am M\"u{}hlenberg 1, DE-14476 Potsdam, Germany  
}
\\[1cm]
 \texttt{lorenzo.casarin@\{itp.uni-hannover.de, aei.mpg.de\}} 

\texttt{tommaso.bertolini@fysik.su.se}
\end{center}
\vspace{0.9cm}
%
\begin{abstract}
\noindent  
We analyse the proposal of defining the Weyl anomaly for classically non-conformal theories as   \(g^{mn}  \langle T_{mn}\rangle - \langle g^{mn} T_{mn} \rangle \), originally put forward by M.\ Duff, 
in the case of a scalar field with quartic self-interaction in 4d. We work in the context of dimensional regularization in curved background to two-loops (first order in the coupling).
We review the original regularized but not renormalized prescription and its ambiguities;  we argue that it cannot be extended to the interacting theory as it fails to provide a finite result.
We then propose an alternative prescription via renormalized 
 expectation values. At one-loop  our candidate reproduces the local heat kernel result, while its extension to interacting theories contains non-local contributions.

 
\end{abstract}
 
\newpage


\newpage

\tableofcontents

\section{Introduction}
Since their discovery by Capper and Duff in \cite{Capper:1973mv,Capper:1974ed,Capper:1974ic}, Weyl anomalies have been a central topic in quantum field theory, see \cite{Duff:1980qv,Duff:1993wm,Birrell:1982ix,Duff:2020dqb} for a number of references and general reviews. These anomalies parametrize the trace of the energy momentum tensor induced by quantum corrections for classically Weyl invariant theories, and provide strong constraints as well as powerful ordering principles in the space of quantum field theories, such as the celebrated \(c\)-theorem in two dimensions \cite{Zamolodchikov:1986gt} and the  \(a\)-theorem in four \cite{Komargodski:2011vj}. 

The Weyl anomaly shares many similarities with the chiral anomaly, but also important differences. The latter manifests itself in a nonzero divergence for the axial current (thereby spoiling its conservation), it is topological and one-loop exact.   
Importantly for the scope of the present paper, when the axial symmetry is explicitly broken by the addition of a mass term, the  divergence of the current is simply the sum of an explicit breaking contribution (proportional to the mass) and of the anomalous term. The Weyl anomaly comes in two types  \cite{Deser:1993yx}: one topological and one built from the Weyl tensor. This second  type of contributions is in general coupling-dependent, although explicitly studying this effect  is difficult, since in perturbation theory  the underlying Weyl symmetry is generically broken by beta functions.\footnote{An explicit example, albeit somewhat exotic, of a langrangian model with type-B anomaly coefficients with explicit coupling dependence is given by the \(6d\) four-derivative vector discussed in  \cite{Casarin:2023ifl}.}  

Furthermore,   the  quantum trace of the stress tensor for a generic QFT, when the Weyl symmetry is explicitly broken  is much less understood than the CFT case.  A better understanding of quantum contributions to the stress tensor has potential applications to QFT in the presence of gravity and in the context of cosmology. The anomalous trace of the stress tensor is also important   in the study of the RG, since conformal symmetry is broken  along the flow. Ambitiously, an interpolating function for the anomaly coefficients can be found and provide insights for the strong \(a\) theorem in four dimensions or its attempted generalisation in six \cite{Elvang:2012st}.  Furthermore, 
a cancellation of some would-be anomaly coefficients has been observed in  \cite{Meissner:2016onk,Meissner:2017qwm}  in the case of certain
Poincaré supergravities. This cancellation has not yet been explained, and is somewhat mysterious since the graviton and the gravitino do not possess two-derivative classically Weyl invariant actions.\footnote{Classically Weyl invariant theories of gravity and supergravity typically involve higher derivative fields. For those,  Weyl anomalies are well-defined, at least at one loop, see  \cite{Fradkin:1985am,Pang:2012rd,Aros:2019tjw,Casarin:2024qdn} and references therein.}

In \cite{Duff:1993wm,Duff:2020dqb} Duff proposed, following the structure of chiral anomalies, to identify the quantum breaking of the Weyl symmetry as
\begin{equation}\label{aaa}
\mathcal A =
 g^{(4) \, mn} \braket{ T_{mn} }_{\text{reg}} - \braket{g^{mn} T_{mn}}_{\text{reg}} 
 \,,
\end{equation}
where the expectation values are taken in the regularized  but not renormalized theory. The reason behind this definition  \cite{Duff:1993wm,Duff:2020dqb} is that the anomaly should be a physical (measurable) quantity and therefore independent of the renormalization  prescription. It should capture the purely quantum contribution   to the stress tensor trace, and for this reason \eqref{aaa}  is usually referred to as Weyl anomaly, although its interpretation is less clear. Notice that for standard classically Weyl-invariant theories the second term in \eqref{aaa} vanishes and the definition reduces to the one used in the original works  \cite{Capper:1973mv,Capper:1974ed,Capper:1974ic} in the context of dimensional regularization. In this work we will refer to $\An$ and its alternative prescriptions as anomalies, even when it is understood that the Weyl symmetry is already broken at the classical level.

An efficient way of computing the  anomaly proper is via the heat kernel (HK) expansion, which retains manifest covariance with respect to the geometry. In this case the anomaly is identified with the HK coefficient of  the kinetic operator \(\Delta\), so that for a conformal scalar\footnote{The generalisation to the case of multiple fields or different spin is immediate, see e.g.\ \cite{Birrell:1982ix}.}
\begin{equation}\label{ahkd} 
g^{mn}\braket{T_{mn}}
=
a_4(\Delta)  \,.
\end{equation} 
The identification of \eqref{aaa} with the heat kernel coefficient is often assumed also to the case in which there is explicit breaking of Weyl symmetry, see e.g.\ \cite{Birrell:1982ix,Duff:1993wm,Meissner:2016onk,Meissner:2017qwm},
\begin{equation}\label{csc}
\mathcal A_\text{hk} = a_4(\Delta)\,.
\end{equation}
It is however a priori not clear which diagrammatic expression it corresponds to and how it extends beyond quadratic (free) level.  For a free scalar with generic (non-Weyl invariant) curvature coupling  \(\Delta = -\Box + \xi R\), the heat kernel prescription gives \cite{Birrell:1982ix,Vassilevich:2003xt}
\begin{equation}\label{ahk} 
\mathcal A _{\text{hk}}  =
 a_4 (-\Box + \xi R)
 = 
	\frac{1}{180(4\pi)^2}\left[
		-\frac12 \EE  
		+ 6\left(1-5 \xi  \right) \Box R
		+ \frac32 \WW
		+\frac52 (6\xi-1)^2 R^2
	\right],
\end{equation}
which features the appearance of an \(R^2\) term, absent in the anomaly proper and showing that this quantity cannot be obtained from functional differentiation.

The definition \eqref{aaa} was studied in dimensional regularization in \cite{Casarin:2018odz}. It was explicitly discussed that the definition is finite and local but presents an ambiguity on the nature of the subtraction term that can be represented by writing explicitly
\begin{equation}\label{aaD}
\mathcal A^{(D)}_\text{reg} =
g^{(4) \, mn} \braket{ T_{mn} }_{\varepsilon} - \braket{g^{(D)\, mn} T_{mn}}_{\varepsilon}  
\qquad\quad (\varepsilon\to0)
\,.
\end{equation}
Indeed, one can subtract the trace in \(D=4\) or \(D= 4-2\varepsilon\) dimensions. In particular,  \cite{Casarin:2018odz} focussed on the case of a  free scalar field with generic curvature coupling was analysed. We added the subscript `reg' to emphasize that it is built of regularized quantities. After that, \cite{Larue:2023qxw} proposed an all-loop modification of \(\mathcal A\) based on dimensional regularization, which effectively extends the prescription \(\mathcal A^{(4-2\varepsilon)}\) to the interacting case.\footnote{Another perspective on  Weyl anomalies for non-conformal theories is  given in 
\cite{Ferrero:2023unz}.} 

In this paper we focus on the prototypical  example of a QFT that breaks  Weyl symmetry explicitly, namely a  scalar in four dimensions with quartic self-interaction.
The  breaking of Weyl symmetry is achieved via the non-conformal quadratic coupling with the curvature    as well as by  nonzero beta functions.
We argue that \(\An_\text{reg}\) as in \eqref{aaa} (or rather the concrete prescriptions  \(\An^{(D)}_\text{reg}\) \eqref{aaD}) does not extend beyond  free level.   We thus modify the prescription \eqref{aaa} by promoting the expectation values   to renormalized (finite) ones, and consider
\begin{equation}\label{aac}
{\mathcal A}_\text{ren} =
 g^{ mn} \braket{ [T_{mn}] } - \braket{[\Theta]}
 \,,
 \qquad\qquad
 \Theta = g^{mn} T_{mn}\,,
\end{equation}
where \(\Theta\) is the four-dimensional  trace of the  classical energy-momentum tensor and the square brackets in the expectation values denote the renormalized composite operators. We construct these renormalized operators in  dimensional regularization  following the well-established tradition of \cite{Brown:1980qq,Brown:1992db,Collins:1984xc,Hathrell:1981zb} and references therein.
In particular, we work in perturbation theory  to first order in the coupling with a formal expansion around a flat background  \(g_{mn} = \delta_{mn} + h_{mn}\) and focus on the contributions to ${ \mathcal{A}}_\text{ren} $ of first and second order in \(h\). We evaluate the former fully, while the latter suffer from the complication of three-propagator subdiagrams which need to be expanded to a nontrivial order in \(\varepsilon\). 
We show that, at free (one-loop) level, \eqref{aac} provides a local result  that  reproduces the heat kernel prescription \eqref{ahk}. 
To circumvent the technical difficulties in evaluating \eqref{aac} to the first order in the coupling at order \(h^2\), we consider the spacetime integral of \(\An_{\text{ren}}\). This is the generally covariantized analogue of setting the momentum of the stress tensor to zero, thereby reducing the integrals to two-propagator diagrams. At two loops we obtain a result that is non-local, and we argue that this is indeed expected in the general case.

In our calculations all the nonlocalities and  departures from the anomaly proper  disappear at the conformal value of the curvature coupling \(\xi=\frac16\), thus the construction might look in this case artificial. However,  this value is  not stable under quantum corrections, which induce an RG flow for this parameter away from the conformal point \cite{Freedman:1974ze,Toms:1982af}. Despite these effects being relevant at a higher order than the ones considered in this paper, our setting is therefore generic.

The   paper is organized as follows. In Section~\ref{sect:set} we give a general review of the formal setting: action, regularization and renormalization in curved background  in perturbation theory.
In Section~\ref{sect:reg} we review the regularized calculations of \cite{Casarin:2018odz}.
 In Section~\ref{sect:ren} we construct the renormalized anomaly \eqref{aac}, commenting on its non-local structure and the two-loop result to first order in the interaction.
Section~\ref{sect:concl} concludes with a summary, a comparison with recent literature, and outlook.     Appendix~\ref{app:not} summarizes notation and conventions;
Appendix~\ref{app:vert} reports lengthy formulae for Feynman vertices;
Appendix~\ref{app:renact} discusses some aspects of the renormalization of the action in curved spacetime that are relevant for our discussion.

\section{Setting and notational remarks}
\label{sect:set}

We consider the scalar action in \(D\) dimensions\footnote{We use lowercase \(d\)  to denote the dimensionally regularized value \(d= 4 - 2 \varepsilon\). We introduce an auxiliary dimension \(D\) to be able to distinguish the two different cases \(D=4\) and \(D=d \) more explicitly.} with a quartic self-interaction and to a geometrical background given by
\begin{equation}\label{act}
S_\varphi = \int\! d^D{x}\sqrt{g}  
		\left[ 
			\frac 12\left(\cd_m \varphi \cd^m \varphi + \xi R \varphi^2 \right)
			+ \frac{\lambda}{4!} \varphi^4
		\right],
\end{equation}
where \(\xi\) is the dimensionless curvature coupling  and \(\lambda\) is classically dimensionless  only in \(D=4\). Weyl invariance of the kinetic term is achieved for  \(\xi=\xi_D := \frac{1}{4} \frac{D-2}{D-1}\).

We note  the equation-of-motion operator 
\begin{equation}\label{bae}
E_\varphi =
\varphi \frac{\delta}{\delta \varphi} S
= \varphi (-\Box + \xi R)\varphi + \frac{\lambda}{3!} \varphi^4  \; ,
\end{equation}
and the stress-tensor   and its \(D\)-dimensional trace are
\begin{equation}\label{baf}
\begin{aligned} 
	T^\varphi_{mn}
& 
	= 	\cd_m \varphi \, \cd_n \varphi
		- \frac{1}{2} g_{mn}\, \cd_a \varphi \, \cd^a \varphi
		-  g_{mn}  \frac{\lambda}{  4!} \varphi^4
\\
& \qquad\qquad{}		+ \xi \, \varphi^2 \left( R_{mn } - \frac{1}{2} g_{mn} R \right)
		- \xi \left( \nabla_m \cd_n \varphi^2 - g_{mn} \Box \varphi^2 \right),
		\\
\Theta^{(D)}
&=
g^{(D) mn}
T_{mn}^\varphi
	= (D-1)( \xi - \xi_D) \Box \varphi^2 - \frac{D-2}{2} E[\varphi]  + \frac{(D-4)}{4!} \lambda\varphi^4 .
\end{aligned}
\end{equation}
The latter indeed shows that the stress tensor is  classically traceless on shell for \(\xi = \xi_D \) at \(D=4\) or when \(\lambda=0\).    In particular we note the value of the classical trace in \(D=4\) dimensions, %
\begin{equation}\label{bas}
\begin{aligned}
\Theta &\equiv \Theta^{(4)}
	= 3\left( \xi - \frac16\right) \Box \varphi^2 
	-E_\varphi.
\\
\end{aligned} 
\end{equation}

As we are going to review in the next subsection,    the equation of motion operator has vanishing expectation value in  dimensional regularization, both in the bare and in the renormalized theory. Since we will be only considering such one-point functions,  we will often  drop it.

\subsection{Regularization}
 We adopt the framework of dimensional regularization with \(d=4-2\varepsilon\), which is standard in both flat and curved spacetime \cite{Birrell:1982ix,Brown:1980qq,Hathrell:1981zb,Collins:1984xc,Brown:1992db,Kleinert:2001ax}.
 For simplicity and ease of exposition we understand the energy scale \(\mu\) and reinstate it only in final expressions. 
 
We are interested in regularized and then renormalized expectation values of \(T^\varphi_{mn}\) and \(\Theta^{(D)}\) which we compute via the path integral
\begin{equation}\label{dcs}
\braket{T_{mn}^\varphi}_\varepsilon = \int \DD{\varphi} e^{-S} T^\varphi_{mn} \,,
\qquad
\braket{\Theta^{(D)}}_\varepsilon  = \int \DD{\varphi} e^{-S } \Theta^{(D)} \,,
\qquad
\int \DD{\varphi} e^{-S }    = 1 \,,
\end{equation}
where the subscript $\varepsilon$ indicates the use of bare dimensionally-regularized correlators.
 Fundamental in our discussion is the observation that 
\begin{equation}\label{khd}
\braket{  \Theta^{(d)}  }_\varepsilon  = \braket{ g^{(d) \, mn} T^\varphi_{ mn } }_\varepsilon 
= g^{(d) \, mn}\braket{T^\varphi_{ mn } }_\varepsilon \,,
\end{equation}
namely when considering the expectation value of the \(D=d\) dimensional trace \(\Theta^{(D=d)}\), the contraction with the metric can be equivalently  taken before or after path integration (or equivalently  before and after expanding in \(\varepsilon\)).
This is possible because for regularized correlators the rule \( g^{(d) \, mn} g^{(d)}_{  mn} = d=4-2\varepsilon \) is valid  inside and outside the path integral symbol. We stress that this holds true only for \(\braket{  \Theta^{(D=d)}  }_\varepsilon \). For \(\braket{  \Theta^{(D=4)}  }_\varepsilon \) it is not the case.

Another important feature of dimensional regularization is that    \(\braket{E[\varphi]} =0 \), since 
\begin{equation}\label{bap}
\braket{E[\varphi]} = \int \DD{\varphi} e^{-S}\varphi(x) \frac{\delta}{\delta \varphi(x)} S
= - \int \DD{\varphi}\frac{\delta}{\delta \varphi(x)} \left(  \varphi(x) e^{-S} \right)
=0\,,
\end{equation}
which vanishes as a functional boundary term.%
\footnote{
In \eqref{bap} we used the standard value  \(\frac{\delta}{\delta\varphi(x)} \varphi(x) = \delta[x-x]=0\) of dimensional regularization \cite{Kleinert:2001ax,Brown:1992db,Hathrell:1981zb,Collins:1984xc}. 
}
 
We wish to evaluate the correlators in \eqref{dcs} in perturbation theory in \(\lambda\). To be able to use the    well-developed diagrammatic technology,  we perform a formal expansion on a flat background  \(g_{mn}=\delta_{mn}+h_{mn}\) and work order by order in  \(h_{mn}\).  In particular, we will need
\begin{equation}\label{baa}
\begin{gathered}
S_\varphi =  
S^{(0)}_{\varphi^2} + S^{(1)}_{\varphi^2} + S^{(2)}_{\varphi^2} + \ldots +
S_{\varphi^4}^{(0)} +S_{\varphi^4}^{(1)}  +\ldots,
\\
T^{\varphi}_{mn}  = 
T  ^{{\varphi^2}(0)} _{mn}
 + T ^{{\varphi^2}(1)} _{mn}
 + \ldots
  + T ^{{\varphi^4}(0)} _{mn}
  + T ^{\varphi^4(1)}_{mn} + \ldots,
\qquad\quad
\Theta    = 
\Theta^{{\varphi^2}(0)}
+
\Theta ^{{\varphi^2}(1)}  
+\ldots,
\end{gathered}
\end{equation}
where the superscript \((n)\) indicates the power of \(h\), and \(\varphi^2,\varphi^4\)  distinguish the free vs.\ interaction contributions.
In particular, \(S^{(0)}_{\varphi^2}+S^{(0)}_{\varphi^4} \) is the flat-space scalar action,\footnote{
We write all indices lowered to emphasize the contraction with the flat metric.
}
\begin{equation}\label{bab}
S^{(0)}_{\varphi^2} 
= \frac12\int \! d^dx \, \partial_m \varphi \ \partial_m \varphi 
\,, 
\qquad
S^{(0)}_{\varphi^4} 
= \frac{\lambda}{4!} \int \! d^d x \, \varphi^4 
\end{equation}
and we adopt the following notation for the   interactions with the background metric
\begin{equation}\label{bba}
\begin{aligned}
S^{(1)}_{\varphi^2}  & =  \int\dm{p}\dm{q}\dm{\ell}
				\ (2\pi)^{d}\delta[p+q+\ell ] \
			  \varphi(p) \varphi(q) h_{mn}(\ell) \ 
			  V^{\varphi^2(1)}_{mn}(p,q,\ell) 
\,,
\\ 
S^{(2)}_{\varphi^2}  &=  \int \dm{p}\dm{q}\dm{k}\dm{\ell}
				\ (2\pi)^{d}\delta[p+q+k+\ell ] \
			  \varphi(p) \varphi(q) h_{mn}(\ell)h_{rs}(k) \ 
			  V^{\varphi^2(2)}_{mnrs}(p,q,\ell,k) \,,
\end{aligned}
\end{equation}
and so on analogously for all terms $S^{(n)}_{\varphi^m}$.
Similarly, for the stress tensor we write\footnote{The vertex functions for \(T_{mn}\) and \(\Theta ^{(D)}\) do not %
 involve a momentum-conserving delta function, as they are  external vertices  injecting momentum in the graph.}
\begin{equation}\label{bcb}
\begin{aligned}
 T^{\varphi^2(0)}  _{mn}
&=  \int \! \dm p \dm q 
				\ e^{i(p+q)x }
			  \varphi(p) \varphi(q)  \ 
			  W^{\varphi^2(0)}_{mn}(p,q) 
  \,,  
\\
T^{\varphi^2(1)}  _{mn}
&=  \int  \! \dm p \dm q \dm \ell 
	  				\ e^{i(p+q+\ell)x }
		  \varphi(p) \varphi(q) h_{ac}(\ell) \ 
		  W^{\varphi^2(1)}_{mnac}(p,q,\ell) ,
\end{aligned}
\end{equation}
and so on.\footnote{In general one   needs to introduce Feynman rules also for \(\Theta\), but in this particular example it is not necessary.}
Explicit expressions for the relevant vertices are reported in Appendix~\ref{app:vert}.

\subsection{Renormalization}\label{sect:defreg}
Renormalizing the theory  on  curved geometry requires the  familiar infinite rescaling of the parameters in the action \eqref{act}   as well as additional    terms to cancel purely gravitational infinities. One therefore considers the total action
\begin{equation}\label{bac}
\begin{aligned}
S &= S_\varphi + S_\gr \,,
\qquad\quad 
S_\gr = \int \! d^d{x} \sqrt{g}  
		\left[ 
		- \alpha \EE
		+ \gamma \WW
		+ \rho R^2
		\right] .
\end{aligned}
\end{equation}
The gravitational contribution is quadratic in the curvature and contains the Euler density, the square of the Weyl tensor and the square of the Ricci scalar (explicit expressions in Appendix~\ref{app:not}).
We have	 an expansion in \(h_{mn}\)  analogous to \eqref{baa},
\begin{equation}\label{baa2}
\begin{gathered}
S_\gr = S^{(2)} + S^{(3)} + \ldots\,,
\qquad
S^{(2)}   =  \int\dm{p}
			   h_{mn}(p)h_{rs}(-p)   \ 
			  V^{ (2)}_{mn,rs }(p) \,,
			  \\
S^{(3)}   =  \int\dm{p}\dm{q}\dm{k}
				(2 \pi)^d \,\delta[p+q+k] \ 
			   h_{mn}(p) \, h_{rs}(q) \,  h_{ac}(k)    \ 
			  V^{ (3)}_{mn rs ac }(p,q,k) \,.
\end{gathered}
\end{equation} 
A finite theory is then obtained by setting
\begin{equation}\label{cfd}
\begin{gathered}
\lambda \to \lambda + \sum_{i\geq 1} \frac{\lambda^{(i)} }{\varepsilon^i} 
\,,
\qquad
\xi \to \xi + \sum_{i\geq 1} \frac{\xi^{(i)} }{\varepsilon^i} 
\,,
\qquad
\varphi \to \left(1 + \sum_{i\geq 1} \frac{z^{(i)} }{\varepsilon^i}   \right)  \varphi 
\,,
\\
\alpha \to  \sum_{i\geq 1} \frac{\alpha^{(i)} }{\varepsilon^i} 
\,,
\qquad\qquad
\gamma \to   \sum_{i\geq 1} \frac{\gamma^{(i)} }{\varepsilon^i} 
\,,
\qquad\qquad
\rho \to    \sum_{i\geq 1} \frac{\rho^{(i)} }{\varepsilon^i} \,,
\end{gathered}
\end{equation}
We work in minimal subtraction  scheme so that  so that the Weyl tensor  is intended as the four-dimensional one and \(\alpha,\gamma,\rho\) are only poles.	 The values of the counterterms in \eqref{cfd} have been computed in the literature long ago and we will quote the relevant ones momentarily. In Appendix~\ref{app:renact} we comment on some aspects of their calculation in the spirit of the present work.

We will need the renormalized stress tensor \([T_{mn}]\) and the renormalized stress tensor trace \([\Theta]\). The square brackets denote renormalized composite operators. Constructing renormalized composite operators in dimensional regularization is  standard also in the curved background context, see  \cite{Brown:1980qq,Hathrell:1981zb,Collins:1984xc,Brown:1992db,Kleinert:2001ax}. Here we summarize the relevant results.

A finite (renormalized) stress-tensor operator is   obtained by differentiation  of  the renormalized full action \eqref{bac} with the renormalized values \eqref{cfd}, so that\footnote{%
An operator is renormalized by requiring that its insertion produces finite correlators. Given any finite correlator, an additional stress-tensor insertion is realized by differentiation with respect to the metric thus without introducing any additional divergence.
}
\begin{equation}\label{dsd}
	[T_{mn}] =-\frac{2}{\sqrt{g}} \frac{\delta}{\delta g^{mn}} [S]
	= -\frac{2}{\sqrt{g}} \frac{\delta}{\delta g^{mn}} [ S_\varphi + S_{\text{grav}} ]
	= 
	 [T^\varphi_{mn}+T^\text{grav}_{mn} ]\,,
	 \qquad
	 	T^\text{grav}_{mn} = -\frac{2}{\sqrt{g}} \frac{\delta}{\delta g^{mn}} S_\text{grav} \,.
\end{equation}
The expansion \eqref{baa} is therefore complemented by
\begin{equation}\label{dsx}
\begin{gathered}
T^\text{grav}_{mn} =  T^{0(1)}  _{mn}
+  T^{0(2)}  _{mn}
+ \ldots\,,
\qquad
\qquad
 T^{0(1)}  _{mn}
=  \int  \dm \ell  
				\ e^{i \ell x}
			   h_{ac}(\ell) \ 
			  W^{0(1)}_{mnac}(\ell) \,,
\\
 T^{0(2)}  _{mn}
=  \int  \dm \ell \dm k   
				\ e^{i (\ell + k) x}
			   h_{ac}(\ell)h_{rs}(k) \ 
			  W^{0(1)}_{mnacrs}(\ell,k)  
\end{gathered}
\end{equation}
Following \eqref{bac} and \eqref{cfd}, these terms are pure poles and are responsible for the anomaly proper.

To construct\footnote{Insertions of \(\Theta^{(d)}\) in arbitrary correlators are produced by differentiation with respect to the conformal factor of the metric, so \(\Theta^{(d)}\) does not require additional subtractions, consistently with \eqref{khd}. In contract, no shortcut is available for \(\Theta^{(4)}\).} a finite operator associated to the four dimensional stress-tensor trace \(\Theta\), we start with the renormalized operator \([\varphi^2(x)] \), which is given by
\begin{equation}\label{ndpp}
[\varphi^2] = Z_2 \varphi_{0} ^2 + Z_{g} R\,,
\qquad
Z_2 = 1 + \sum_{i\geq 1} \frac{z_2^{(i)}}{\varepsilon^i} 
\,,
\qquad
Z_{g} = \sum_{i\geq 1} \frac{z_{g}^{(i)}}{\varepsilon^i} 
\,.
\end{equation}  
We then define the renormalized operator associated to \eqref{bas} as\footnote{The equation of motion operator does not require any additional subtraction \cite{Brown:1980qq,Brown:1992db,Hathrell:1981zb} and it has vanishing expectation value.}
\begin{equation}\label{ndp}
[\Theta] = 3 \left(\xi-\frac16\right) \Box [\varphi^2] - E_\varphi\,,
\end{equation}
where \(\xi\) is the renormalized (finite) value. 
 
The couterterms \eqref{cfd} have been computed in the literature with a combination of diagrammatic and heat kernel methods.  To the relevant order the counterterms are     \cite{Toms:1982af,Brown:1992db,Brown:1980qq}
\begin{equation}\label{wsf}
\begin{gathered}
\xi^{(1)} =  
\frac{6 \xi -1}{12 (4\pi)^2} \lambda 
\,,
\qquad
\xi^{(2)} =  0
\,,
\qquad
\alpha^{(1)} = \frac{1}{720 (4 \pi)^2} 
\,,
\qquad
\alpha^{(2)} = 0
\,,
\\
\gamma^{(1)} = \frac{1}{240 (4 \pi)^2} 
\,,
\qquad
\gamma^{(2)} = 0
\,,
\qquad
\rho^{(1)} = 
 \frac{(6 \xi -1)^2}{144 (4\pi)^4}   
\,,
\qquad
\rho^{(2)} = 
 - \frac{(6 \xi -1)^2}{288 (4\pi)^2} \lambda 
\,,
\\
\lambda^{(1)} = 0 
\,, 
\qquad
z^{(1)} = 0 %
\,, 
\qquad
z_2^{(1)} =
\frac{\lambda}{2 (4 \pi)^2} 
\,,
\qquad
z_g^{(1)} = 
\frac{6 \xi -1}{6 (4 \pi)^2} \,.
\end{gathered}
\end{equation}
In particular,  at the order in which we are working there is no renormalization of the coupling \(\lambda\) nor there is wavefunction renormalization.  
 
\section{The regularized expression}
\label{sect:reg}

\subsection{Ambiguity and one loop (free theory) results}
In this subsection we review the one-loop calculation of  \cite{Casarin:2018odz} (cf.\ also \cite{Casarin:2021fgd}) and we extend some of the results and discussions.

In the context of dimensional regularization we can interpret \eqref{aaa} in different ways, i.e.\ there is an intrinsic ambiguity.
\begin{equation}\label{odj}
\mathcal A^{(D)}_\text{reg} =
 g^{(4)\, mn} \braket{ T_{mn} }_{\varepsilon} - \braket{ \Theta ^{(D)}}_{\varepsilon}
 \qquad\qquad (\varepsilon\to0)
 \,,
\end{equation}
with \(D= 4\) or \(D=d=4-2\varepsilon\).   The origin of this ambiguity can be appreciated by looking at the explict expression \eqref{baf}: the difference between \(\Theta^{(4)}\) and \(\Theta^{4-2\varepsilon}\) is of order \(\varepsilon\) and thus they produce different terms when combined with the poles of loop integrals.

The case \(D=d\) has a computational advantage: one can compute \(\mathcal A^{(d)}_\text{reg}\) with the knowledge of the divergent part of  \(\braket{ T_{mn} } \) only, without the need to consider the more complicated finite pieces. Indeed, as a consequence of \eqref{khd} we can write \eqref{odj} for \(D=d\) as 
\begin{equation}\label{odg}
\mathcal A^{(d)}_\text{reg} = (  g^{(4)\, mn}  -  g^{(d)\, mn}   )   \braket{ T_{mn} } _{\varepsilon} 
 \qquad\qquad (\varepsilon\to0)
 \,.
\end{equation}
This expression  shows two important features. First, only terms in \(\braket{ T_{mn} } _{\varepsilon} \) proportional to the metric \(g_{mn}\) contribute: everything else cancels in the difference,  as e.g.\ \(g^{(D)\, mn} R_{mn} = R    \) for any \(D\). Second, only the pole   of  \( \braket{ T_{mn} } _{\varepsilon} \) contributes, as  can be  seen by writing
\begin{equation}\label{fde}
\braket{T_{mn}}_\varepsilon
= \frac{1}{\varepsilon} (  P_{mn} + g_{mn} Q_{mn}   )
+    F_{mn} + \mathcal{O}(\varepsilon)\,,
\end{equation}
where \(P_{mn}\) denotes tensor structures that are not proportional to \(g_{mn}\). 
From \eqref{odg} we thus have
\begin{equation}\label{cxa}
\mathcal A^{(d)}_\text{reg} =  
\lim_{\varepsilon \to 0 }\left[
\frac{4}{\varepsilon} Q_m^m
-   
\frac{4-2\varepsilon}{\varepsilon} Q_m^m
\right]
=2 Q_m^m\,,
\end{equation}
where \(P_{mn}\) and \(F_{mn}\) have dropped since \(  g^{(4)\, mn} P_{mn}  =  g^{(d)\, mn} P_{mn} \) and  \(  g^{(4)\, mn} F_{mn}  =  g^{(d)\, mn} F_{mn} + \mathcal O (\varepsilon)\). We notice that this argument does not   rely on perturbative expansion in \(h\): if the full covariant expression for the (local) pole of \(\braket{T_{mn}}_\varepsilon\) is known (as is e.g.\ using the heat kernel expansion), 	this immediately gives the covariant result.

\begin{figure}[htb]
\centering 
\includegraphics[scale=0.5]{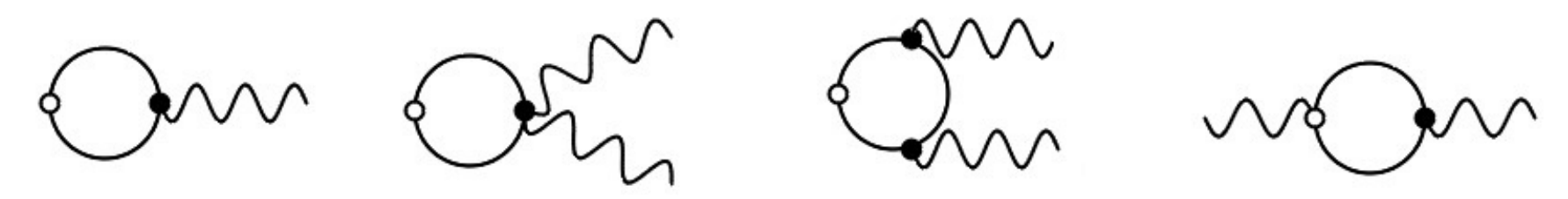}  
\caption{Diagrammatic representation of \(\bb{ T_{mn}(x) } _{\varepsilon\, \mathcal O(h^1)} + \bb{ T_{mn}(x) } _{\varepsilon\, \mathcal O(h^2)}  \)  in \eqref{fks}. Black dots denote interaction vertices \(V\), white dots denote stress-energy tensor vertices  $W$. Solid lines represent scalar propagators and  wavy lines  the metric perturbation $h$.}
\label{fig:feyn1}
\end{figure}

To perform this calculation diagrammatically we expand in \(g_{mn} = \delta_{mn} + h_{mn}\). Using the definitions of section~\ref{sect:defreg} we  have to consider the following terms\footnote{We   denote by $\bb{...}_\varepsilon$ the expectation value   taken with respect to the  flat-space free  theory.  }
\begin{equation}\label{fks}
 \braket{ T_{mn}(x) } _{\varepsilon} 
 =  %
 -  \bb{ T^{(0)}_{mn} S^{(1)} _{\varphi^2}}  _{\varepsilon}  
    -  \bb{ T^{(0)}_{mn} S^{(2)} _{\varphi^2}} _{\varepsilon}  
  +\frac12  \bb{ T^{(0)}_{mn} S^{(1)} _{\varphi^2}S^{(1)} _{\varphi^2}} _{\varepsilon}  
   -  \bb{ T^{(1)}_{mn} S^{(1)} _{\varphi^2}} _{\varepsilon}  
   + \mathcal O (h^3)
 \,,
\end{equation}
whose diagrammatic representation is pictured in figure~\ref{fig:feyn1}. 
The associated integrals are listed in Appendix \ref{app:vert}.  
They can be expanded   in \(\varepsilon\)   and  the anomaly can be obtained \(\An^{(d)}\) by direct application of \eqref{odg} (i.e.\ \eqref{cxa}). Owing to locality,  the full \(g\) dependence can be reconstructed  by demanding general covariance. We refer to  \cite{Casarin:2018odz,Casarin:2021fgd} for  details and we simply state the result,
\begin{equation}\label{fsd}
\begin{aligned}
\mathcal A^{(d)}_\text{reg}
&=
	\frac{1}{180(4\pi)^2}\left[
		-\frac12 \EE  
		+ 6\left(1-10(6\xi-1)^2\right) \Box R
		+ \frac32 \WW
		+\frac52 (6\xi-1)^2 R^2
	\right].
\end{aligned}
\end{equation}

The prescription \(\mathcal A ^{(4)}_\text{reg}\) was only briefly described in \cite{Casarin:2018odz,Casarin:2021fgd} and here we provide some more detail. At  first sight, this case  seems to require the full evaluation of the finite parts of the correlators. However we notice the seemingly trivial rewriting
\begin{equation}\label{dsw}
\begin{aligned}
\mathcal A^{(4)}_\text{reg} & = 
 g^{(4)\, mn} \braket{ T_{mn} }_{\varepsilon} - \braket{ \Theta ^{(d)}}_{\varepsilon}
+\braket{ \Theta ^{(d)}}_{\varepsilon} - \braket{ \Theta ^{(4)}}_{\varepsilon}
 \qquad\qquad (\varepsilon\to0)
 \\
 & = 
 \mathcal A^{(d)}_\text{reg} + \braket \Delta\,,
\end{aligned}
\end{equation}
where  we defined 
\begin{equation}\label{dsv}
\braket \Delta := \braket{ \Theta ^{(d)}}_{\varepsilon} - \braket{ \Theta ^{(4)}}_{\varepsilon}
 \qquad\qquad (\varepsilon\to0).
\end{equation}
It is clear that the splitting in the second line \eqref{dsw} is meaningful, namely that the two terms \( \mathcal A^{(d)}_\text{reg} \) and \(  \braket{\Delta}\) are separately finite: the former is discussed above, the latter follows from
\begin{equation}\label{dss}
\Delta := \Theta ^{(d)} - \Theta ^{(4)}
 = \varepsilon  \cdot\frac{4 \xi - 1 }{2} \Box \varphi^2 + \varepsilon E _\varphi.
\end{equation}
Thus, computing  \(\braket \Delta_\varepsilon\) the second term vanishes (cf.\ \eqref{bas}) and the first one gives a finite and local result, which is proportional to \(\Box R\) on dimensional and covariance grounds. Indeed we find
\(\braket \Delta 
  =
- {15} ( 4 \xi -1 ) (6 \xi -1)   \Box R  \) and 
as a result
\begin{equation}\label{fsd4}
\begin{aligned}
\mathcal A^{(4)}_\text{reg}
&=
	\frac{1}{180(4\pi)^2}\left[
		-\frac12 \EE  
		+ 6\left(1-5 \xi  \right) \Box R
		+ \frac32 \WW
		+\frac52 (6\xi-1)^2 R^2
	\right].
\end{aligned}
\end{equation}

We can see that a generic value of \(\xi\) features the appearance of  \(R^2\) in the anomaly with either prescription. Since this is not compatible with the Wess-Zumino consistency conditions, it follows that the quantity \(\mathcal A_\text{reg}^{(D)}\) is not a functional derivative of an effective action as already noticed in the original paper \cite{Duff:1993wm}. %
This observation was anticipated in the free-scalar calculation of \cite{Casarin:2018odz,Casarin:2021fgd} and is also discussed in \cite{Larue:2023qxw}, where the authors introduce an all-loop definition for the conformal anomaly in dimensional regularization which effectively extends the prescription  \(A^{(d)}_\text{reg} \).  As a consequence,  all four coefficients in \eqref{fsd4} are physical and the difference between the two possible choices, $\mathcal A^{(4)}_\text{reg}$ and $\mathcal A^{(d)}_\text{reg}$, cannot   be reabsorbed by the introduction of counterterms in the action.\footnote{This point is overlooked in \cite{Casarin:2018odz} and  corrected in \cite{Casarin:2021fgd}.}  In fact, finite counterterms cancel between the two terms in \eqref{odj}.
The  ambiguity \(D=4\) vs.\ \(D=d\) in the subtraction \(\mathcal A_\text{reg}^{(D)}\)  is not discussed in \cite{Duff:1993wm} where the quantity $\mathcal A^{}_\text{reg}$ was first introduced, and to our knowledge it is not discussed anywhere else besides the references above. 
Finally, we observe that the heat-kernel identification \eqref{ahk} coincides with the prescription \(\mathcal A^{(4)}_\text{reg}\). This is a nontrivial result that, to our knowledge, was so far discussed only in \cite{Casarin:2018odz,Casarin:2021fgd}.

This concludes our review of the calculation of \cite{Casarin:2018odz}, which hopefully clarifies some incorrect comments reported elsewhere.%
\footnote{
In the conclusions of both arxiv v1, v2 and of the journal version, reference \cite{Ferrero:2023unz} comments that   \(A_\text{reg}^{(d)}\) as in \eqref{fsd}  is ``obtained using dimensional regularization and a perturbative expansion around flat space, together with a dose of intuition to use the right amount of on-shellness'' to simplify the stress-tensor trace. Similar statements appear in the introduction. It should be clear from the discussion above that this remark is incorrect in two ways: i) operatively the result \eqref{fsd} does not directly depend on \(\Theta^{(d)}\) but relies on the epsilon expansion of the %
\(\braket{T_{mn}} _\varepsilon\) only; ii) computing \(\braket {\Theta^{(d)}}_\varepsilon\)   directly neglecting the e.o.m.\ operator \(E\), as done for \(\braket{\Theta}_\varepsilon \) in \(\mathcal{A}_{\text{reg}}^{(4)} \)  is not a problem because \(\braket E=0\) in dimensional regularization as in  \eqref{bap}. 
Unfortunately  the authors of \cite{Ferrero:2023unz} did not share their impression with those of \cite{Casarin:2018odz} prior to publication. 
For completeness, we  note that  \(\mathcal{A}_{\text{reg}}^{(4)} \) is not discussed in  \cite{Ferrero:2023unz}.
}
In the next section we verify our claim that $\mathcal A^{(D)}_\text{reg}$ is divergent in the limit where the regulator is removed in an interacting theory.

\subsection{Failure at two loops (first order in the coupling)}
We will now argue that the definition \eqref{odj} does not  provide a finite quantity at higher loop order by showing an explicit two-loop divergence proportional to \(\Box R\). As discussed at the end of the previous subsection, it is  an un-ambiguous quantity, in contrast to the anomaly proper.

It is enough to consider the term of order \(\mathcal O(h ^1) \). The relevant contribution is
\begin{equation}\label{fjb}
\begin{aligned}
& \braket{T_{mn} (x)}_\varepsilon|_{\mathcal{O}(h^1,\lambda^1)}= 
\bb{T^{(0)}_{mn} (x) \ S^{(1)} _{\varphi^2} \ S^{(0)} _{\varphi^4}} _\varepsilon
\\*
&\qquad{}=
\lambda \int\! d^dx   \int \! \dm{q}   \ e^{iqx}\ h_{rs}(x)
	\int \! \dm{p} \frac{1}{p^2(q-p)^2} 
				W^{\varphi^2(1)}_{mn}(p,q-p,-q)
			\	 V^{\varphi^2(1)}_{rs}(-p,p-q,q) 
\\
& \qquad{}=
-\lambda
\frac{    ( 6 \xi -1 )^2  }{(4\pi)^4 \varepsilon^2} 
\int\! d^dx   \int \! \dm{q}   \ e^{iqx}\ h_{rs}(x)
\frac{  (\delta_{mn} q^2  - q_{m} q_{n})(\delta_{rs} q^2   - q_{r} q_{s})}{72  }
+ \mathcal{O}(\varepsilon^{-1})
\end{aligned}
\end{equation}
A direct calculation on the lines of \eqref{cxa} immediately shows that the presence of a double pole proportional to the metric renders the anomaly \(\An^{(d)}_\text{reg}\) divergent, hence the definition \eqref{odg} is insufficient to accommodate for interactions.%

For ease of exposition we do not discuss the analogous calculation for \(\An^{(4)}_\text{reg}\), as the idea is essentially the same and its divergent nature is implicit in the results of the next sections.

\section{Renormalized construction}
\label{sect:ren}

Having established the insufficiency of the regularized prescription, we turn to the definition \eqref{aac} based on renormalized correlators,
\begin{equation}\label{cfj}
\An_{\text{ren}}=g^{ mn} \braket{ [T_{mn}] } - \braket{[\Theta]}
 ,
\end{equation}
which by construction works to arbitrary loop order and does not have any ambiguity once a renormalization scheme is chosen. We work in minimal subtraction.

Let us see the consequences of this definition in practice. Here we focus on the \(\mathcal O (h^1,\lambda^0)+\mathcal O (h^1,\lambda^1) \) contribution to parallel  the discussion of the previous section.  As we shall see, we do not need the  more complex \(\mathcal O (h^2,\lambda^0) \) term to fully obtain \(\An_\text{ren}\) in the free case. The contribution \(\mathcal O (h^2,\lambda^1) \) is even more complicated and will be considered in a simplified setting in a later section.

The first term of \eqref{cfj} can be computed from the renormalized effective action on a curved background. Here we work in series of \(h\), so
\begin{align}\label{fdlj}
\Gamma[g]& = 
\int \! \dm{^4p} \dm{^4q} h_{mn} (p) h_{rs} (q)  \, (2\pi)^4 \, \delta[p+q] \, \Gamma_{mnrs}(p,q) 
+ \ldots\,,
\\
\braket{T^{mn}(x)} &= 
- \frac{2}{\sqrt{g(x)}} \frac{\delta}{\delta g_{mn}(x)} \Gamma[g] 
= \int \dm k  e^{ikx} \biggl[ 
-4    h_{rs}(k)  \, \Gamma_{mn rs}(- k,k)
   + \ldots  \biggr] 
\end{align}
from which the trace can be readily computed and expanded in \(h\),
\(g^{mn}\braket{T_{mn}}
=
g_{mn}\braket{T^{mn}} =(\delta_{mn}+h_{mn})\braket{T^{mn}} \).
The second term of \eqref{cfj} is essentially given by the diagrammatic evaluation of \(\braket{[\varphi^2]}\) following  the definition \eqref{dss}.

In particular, to the lowest order in the metric perturbation, we obtain 
\begin{align}\label{lkp} 
   g^{mn} \braket{[T_{mn}]}_{\mathcal{O}(h^1)} & =  
\int \!\dm{p}
   e^{ipx} h_{m n}(p) \cdot  
      ( p_{m} p_{n}-\delta_{mn} p^2 ) p^2 
   \cdot
     \\*[-0.5em] \nonumber
     & \hspace{1cm} \cdot
      \bigg[
         \frac{11-60 \xi + 15 (6 \xi -1)^2 \log \frac{p^2}{\bar{\mu }^{2} } }{180 (4\pi)^2}
         - \lambda\frac{\big[3 (6\xi-1)  \log \frac{p^2}{\bar{\mu }^{2} } -1 \big]^2 }{216(4\pi)^4 }  
          \bigg]
   + \mathcal{O}(\lambda^2) \,,
 \\* \label{lkpp} 
      \braket{[\Theta]}_{\mathcal{O}(h^1)} 
& = 
 \int \!\dm{p}
e^{ipx} h_{m n}(p) \cdot
   ( p_{m} p_{n}-\delta_{mn} p^2 ) p^2 \cdot
     \\*[-0.5em]  \nonumber
     & \hspace{2.5cm} 
     \cdot(6 \xi -1)
     \frac{
     \Big[    2 (4\pi )^2 + \lambda  \log\frac{p^{2} }{\bar{\mu }^{2} }  \Big] \
     \Big[ 3 ( 6 \xi - 1 ) \log\frac{p^{2} }{\bar{\mu }^{2} }  -1  \Big] 
     }{72 (4\pi )^4} 
   + \mathcal{O}(\lambda^2)  \,,
\end{align} 
where \(\bar \mu^2 := \mu^2 e^{\gamma_\text E / 4 \pi} \).
As a result we can recognize the convariant structure
\begin{equation}\label{fdo} \mathcal A_{\text{ren}}
   =  \frac{5 \xi -1}{30 (4\pi)^2}  \Box R  
+  \lambda \frac{
  \Big[ 3 ( 6 \xi - 1 ) \log\frac{\Box }{\bar{\mu }^{2} }  -1  \Big]  \
   \Big[ 6 ( 6 \xi - 1 ) \log\frac{\Box }{\bar{\mu }^{2} }  -1  \Big] 
   }{216 (4\pi )^4}\Box R   + \mathcal{O}(h^2,\lambda^2)\,.
\end{equation}
In \eqref{lkp} and \eqref{lkpp} there are nonlocal terms both in the free and in the interacting contributions. We notice that when \(\lambda=0\) these exactly cancel, while they survive in the interacting case.
The free contribution agrees with the regularized value \(\An^{(4)}_\text{reg}\) in \eqref{fsd4} (and thus with the heat kernel prescription \(\An_\text{hk}\)). In fact, as we shall see in the next section, this agreement  can be argued on general grounds at least in the massless case and we do not need an explicit calculation to obtain in general result 
\begin{equation}\label{fdo0}
\mathcal A_{\text{ren}}
= \An^{(4)}_\text{reg} \qquad\qquad \text{(free theories)}.
\end{equation}
Effectively, this means that the renormalized definition extends the HK prescription to arbitrary loop number.

\subsection{Some properties of the renormalized anomaly}

In this section we make some general consideration on the renormalized  $\An_{\text{ren}}$, comparing it between   free     and   interacting  theory. We focus on massless theories for simplicity.

We begin with the renormalized anomaly of the free theory. Interestingly, it reproduces the result of the regularized prescription \(\mathcal A^{(4)}_\text{reg}\).  In particular, it is local: the nonlocal contributions   cancel in the difference. We can indeed see this result on general grounds.
Denoting by \((0)\) bare  quantities, there are  only one-loop simple-pole geometrical counterterms and  
\begin{equation}\label{fkp}
\begin{gathered}
S= S^{(0)}+ S_{\text{ct}1}\,,
\qquad
[T_{mn}] = T_{mn}^{(0)} -\frac{2}{\sqrt g} \frac{\delta S_{\text{ct}1}}{\delta g^{mn}} \,,
\qquad
T_{mn}^{(0)}= -\frac{2}{\sqrt g} \frac{\delta S^{(0)}}{\delta g^{mn}} \,,
\qquad
\Theta = g^{(4)\, mn} \,T_{mn}^{(0)}\,,
\\
[\Theta]  =   \{ g^{(4)\, mn} \,T_{mn}^{(0)}  \}  + \theta_{\text{ct}\,g}
\,,
\,\qquad
\theta_{\text{ct}\,g} = Z_{\Box R} \Box R + Z_{R^2} R^2 + Z_{W^2} \WW+ Z_{\EE} \EE\,.
\end{gathered}
\end{equation}
The brackets \(\{ ... \}\) denote the renormalized composite operator without the contribution proportional to the identity operator, which is \(\theta_{\text{ct}\,g}\) and contains only  poles. We dropped irrelevant e.o.m.\ terms. In the case of the massless scalar, only \(Z_{\Box R} \neq0\) in minimal subtraction, cf.\ \eqref{ndp}.  
The crucial point, as we are going to see, is that   \( \{ g^{(4)\, mn} \,T_{mn}^{(0)}  \} =   g^{(4)\, mn} \,T_{mn}^{(0)}   \) for free theories, but typically not when interactions are present, cf.\ \eqref{ndp},\eqref{wsf}.

We now  consider
\begin{equation}\label{dsl}
\An_{\text{ren}} = g^{(4)\,mn}\braket{[T_{mn}]} - \braket{[\Theta]}\,.
\end{equation} The indication of the dimension \(g^{(4)\,mn}\) in \(   \An_\text{ren} \) is naturally redundant, as the expressions in the right hand side are finite and renormalized, so they are in \(4\) dimensions, but we keep it for clarity. 
Explicitly  \(\An_\text{ren}\)   becomes
\begin{equation}\label{fkq}
\begin{aligned}
{}
\An_{\text{ren}}
& = 
g^{(4) \, mn}  \braket{T^{(0)}_{mn}} _\varepsilon
- \braket{ g^{(4)\, mn} \,T_{mn}^{(0)} }_\varepsilon
-\frac{2}{\sqrt g} g^{(4) \, mn}   \frac{\delta S_{\text{ct}1}}{\delta g^{mn}}
-  \theta_{\text{ct}\,g}
\qquad\qquad(\varepsilon\to 0)
\\
& =
\An^{(4)}_\text{reg}
- \lim_{\varepsilon\to 0 } \left( 
\frac{2}{\sqrt g} g^{(4) \, mn}   \frac{\delta S_{\text{ct}1}}{\delta g^{mn}}
+ \theta_{\text{ct}\,g} \right)
\end{aligned}
\end{equation}
We have used that the difference of the first two terms in the first line is of order \(\varepsilon\), therefore it produces a finite and local result in the \(\varepsilon\to 0\) limit which is exactly \(\An^{(4)}_\text{reg}\).
The fact that we are still taking the trace in \(4\) dimensions%
\footnote{We could have extended the trace to \(d=4-2\varepsilon\) dimensions. 
In this case \( g^{(4) \, mn}  \braket{ T^{(0)}_{mn}}  \)  is replaced by \( g^{(d) \, mn}    \braket{T^{(0)}_{mn}} = \braket{g^{(d) \, mn} T^{(0)}_{mn}} = \braket{\Theta^{(d)}} \) and \(g^{(d) \, mn}   \frac{\delta S_{\text{ct}1}}{\delta g^{mn}}\) contains in addition to \eqref{csf} a finite piece, which gives rise to the anomaly.
The conclusion is the same.}   implies that, from the counterterms in \(S_{\text{ct}1}\), we only have a divergent contribution proportional to
\begin{equation}\label{csf}
g^{(4) \, mn} \frac{\delta S_{\text{ct}1}}{\delta g^{mn}} 
\sim 
\frac1\varepsilon
g^{(4) \, mn} \frac{\delta  }{\delta g^{mn}} \int \! \sqrt{g} R^2  
\propto
\frac1\varepsilon
\Box R\,.
\end{equation}
The other contributions vanish when taking the \(4\) dimensional trace. By definition,
\(\theta_{\text{ct}\,g}\) contains only poles.  By construction   \(\An_\text{ren}\) is finite, so the divergent pieces must cancel.

This argument relies on the fact that  \( \{ g^{(4)\, mn} \,T_{mn}^{(0)}  \} =   g^{(4)\, mn} \,T_{mn}^{(0)}   \), which is true   at free level. Including   interactions produces additional `wavefunction' renormalization factors that induce new terms in perturbation theory cf.\ \eqref{ndpp},\eqref{ndp}.  In contrast, \([T_{mn}]\) does not require additional subtractions beyond the standard action renormalization of the action (which involves only non-composite operators), so that a cancellation of the nonlocalities in \(\An_\text{ren}\) seems unlikely on general grounds. Our calculation in the scalar model  supports this, cf.\ \eqref{fdo} and the following section.

\subsection{Integrated anomaly}\label{sect:inta}

Computing \(\An_\text{ren}\) to two loops and third order in the metric perturbation \(h\) requires considering the \(\varepsilon\) expansion to high order of integrals of formidable complexity. To simplify the problem we consider the integrated quantity
\begin{equation}\label{dsf}
\mathbb A  =
\int \!\dd{^4x}\sqrt{g(x)}\ \An_{\text{ren}}(x)
=
\int \!\dd{^4x}\sqrt{g(x)}\ g^{(4)\,mn}(x)   \braket{[T_{mn}](x)}
\,,
\end{equation}
which remarkably, in the present example, does not depend on \(\braket{[\Theta]}\) since  that is a total derivative. The correlator is the finite, renormalized one, and we emphasized that the trace is taken in $D=4$ for additional clarity.

\begin{figure}[htb]
\centering 
\includegraphics[scale=0.4]{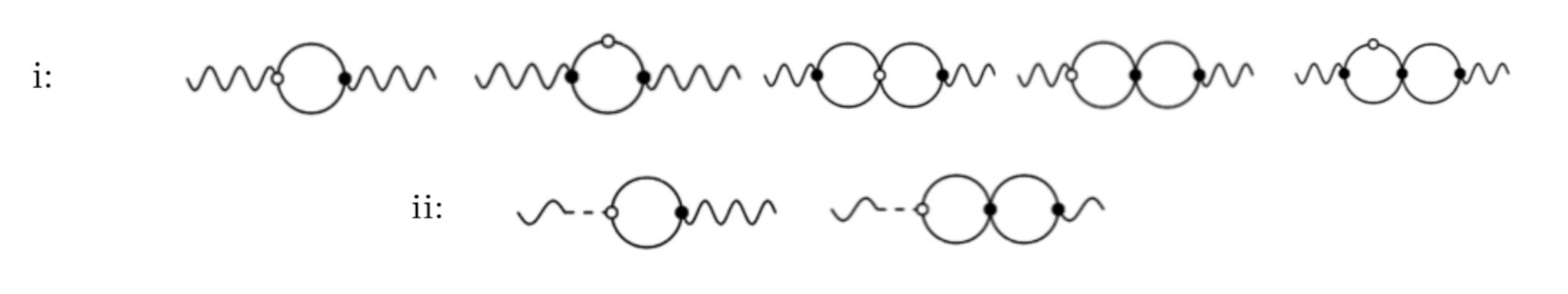}  
\caption{Diagrammatic representation of the bare Feynman integrals  in \eqref{dcl}. Black dots are vertices \(V\) from the action, the white dot is a vertex \(W\) from the stress tensor. Wavy lines are metric perturbations \(h\).   Metric perturbations  with  a dashed line represent the external  $h_{rs}$ that does not come from the correlator.  }
\label{fig:feyn2}
\end{figure}

We focus here on the \( \mathcal{O}(h^2)  \)  contribution to \eqref{dsf} 
\begin{equation}\label{svs}
\left(   \mathbb A  \right)_{\mathcal O (h^2)}
=
\delta_{mn} 
\int \dd{^4x}  \braket{[T_{mn}] }_{\mathcal{O}(h^2)}
+
 \left( \tfrac12 \delta_{rs}\delta_{mn} - \delta_{r  (m}  \delta_{n ) s  } \right) 
 \int\dd{^4x} h_{rs} \braket{[T_{mn}]}_{\mathcal{O}(h^1)}
\,.
\end{equation} 
As the expectation values are computed in dimensional regularization, they have the   structure
\begin{equation}\label{fnd}
 \braket{T_{mn}}_{\mathcal{O}(h^n)}
 =
 \lim_{\varepsilon\to 0}\left[
  \braket{T^{(0)}_{mn}}_{\mathcal{O}(h^n)}
  +
    \braket{T^{\text{ct}}_{mn}}_{\mathcal{O}(h^n)}
    \right]	,
\end{equation}
where the second term indicates countertem contributions.
We find it convenient to extend the  dimensionality of the integrals and of all the delta symbols in  \eqref{svs} from \(4\) to \(d=4-2\varepsilon\) dimensions.  This is consistent   because the correlators are already finite, so this choice does not influence the limit \(\varepsilon\to0\). We can thus rewrite \eqref{svs}  as
\begin{equation}\label{dcl}
\begin{aligned}
\left(   \mathbb A  \right)_{\mathcal O (h^2)}
& =
 \lim_{\varepsilon\to 0} \int \dd{^dx} 
 \delta_{mn}  \left[
  \braket{T^{(0)}_{mn}}_{\mathcal{O}(h^2)}
  +
    \braket{T^{\text{ct}}_{mn}}_{\mathcal{O}(h^2)}
    \right]	 
\\
\qquad\qquad
& \qquad\qquad
+
  \lim_{\varepsilon\to 0}\int \dd{^dx}  
  h_{rs}(x)
  \,
   \left( \tfrac12 \delta_{rs}\delta _{mn} - \delta_{r  (m}  \delta_{n ) s  } \right) 
\left[
  \braket{T^{(0)}_{mn}}_{\mathcal{O}(h^1)}
  +
    \braket{T^{\text{ct}}_{mn}}_{\mathcal{O}(h^1)}
    \right]	 .
    \\
    & =
     \lim_{\varepsilon\to 0}
      \delta_{mn}
     \int   \dm k h_{rs}(k) h_{ac}(-k)
      \ T^{\text i}_{mnrsac}(k)
    \\
    \qquad\qquad
    & \qquad\qquad
    + 
     \lim_{\varepsilon\to 0}
     \left( \tfrac12 \delta_{rs}\delta _{mn} - \delta_{r  (m}  \delta_{n ) s  } \right) 
     \int   \dm k h_{rs}(k) h_{ac}(-k) 
      \ T^{\text {ii}}_{mnac}(k)
\end{aligned}
\end{equation}
Proceeding in this way  has the advantage of setting to zero external momenta \emph{before} expanding in  \(\varepsilon\), thus reducing the diagrams to manageable two-propagator integrals and avoiding IR divergent logs.
In the second step we wrote the integrands in momentum space and indeed implemented  the momentum conservation arising from the integration over \(x\). The diagrammatic representation is   in figure~\ref{fig:feyn2} and we refer to Appendix \ref{app:vert} for the expressions of the corresponding integrals.  

The values in \eqref{wsf} of the counterterms make these expressions finite, providing a consistency check.  
As a result we obtain the integrated anomaly \(\mathbb A\)
\begin{equation}\label{fsj}
\begin{aligned}
(\mathbb A)_{\mathcal O (h^2)}
={}
&	\frac{1}{180 (4\pi)^2} 
\int \!\!
		\frac{d^4 k }{(2\pi)^4 } 
	h_{rs}(k) \ h_{ac} (-k) \times  \Big\{ 
\frac{3}{4} \delta_{ar}\delta_{cs} k^4
- \frac 32 \delta_{ar} k^2 k_s k_c  \\
& 
+\left( 
90 \xi^2  -30 \xi +\frac94 
 - \frac 5{6} \frac{\lambda}{(4 \pi)^2} (6 \xi -1)
 \left(1 - 3 (6 \xi -1) \log \frac{k^2}{\mu} \right)
  \right) 
  (\delta_{rs}\delta_{ac}k^4 
  -2 \delta_{rs} k_a k_c k^2
)
\\
& 
+\left( 
90 \xi^2  -30 \xi + 3 
- \frac 5{6} \frac{\lambda}{(4 \pi)^2} (6 \xi -1)
\left(1 - 3 (6 \xi -1) \log \frac{k^2}{\mu} \right)
\right) 
k_ak_ck_rk_s
		\Big\}.
\end{aligned}
\end{equation}
We recognize the  covariant structure
\begin{equation}\label{fdr}
\begin{aligned}
\mathbb A
=
\frac{1}{180 (4 \pi)^2} 
\int \! d^4 x \sqrt{g} \ 
	 \Big\{
& \frac32 \WW
+ (6 \xi -1)
\left(\frac{5}{2} (6 \xi -1)-\frac{5 \lambda  }{6 (4 \pi)^2}\right)R^2
\\
& \qquad 
+ \frac{5 \lambda  (1-6 \xi )^2}{2 (4 \pi)^2}
R \log \frac{\Box}{\bar \mu^2} R
		\Big\}
 +\mathcal O (h^3 , \lambda^2) .
\end{aligned}
\end{equation}
The Euler term \(\EE\) is a total derivative in four dimensions, hence it disappears after integration, as well as the manifest total derivative \(\Box R\).  
 Interestingly, the \(c\) coefficient  is in any case undeformed by \(\lambda\) at first order.\footnote{This carries  resemblance with the analysis of \cite{Bautista:2020bjy,Bautista:2022kmh} based on the analysis of stress-tensor correlators.} 

We see from \eqref{fsj} and \eqref{fdr} that the departure from conformality brings nonlocalities in  \(\An_\text{ren}\) together with an explicit dependence on the energy scale \(\mu\) besides the  implicit one induced by the renormalization of the parameters.

\section{Concluding remarks}
\label{sect:concl}

In this paper we have explored the characterisation of the quantum contributions to nonconformal theories \(\An \) \eqref{aaa} proposed by M.\ Duff in \cite{Duff:1993wm,Duff:2020dqb}. We studied in dimensional regularization  the explicit  example of a scalar field with a generic curvature coupling and a quartic self-interaction.

The free case was studied in \cite{Casarin:2018odz,Casarin:2021fgd} using regularized but not renormalized correlators. In particular, an ambiguity in the definition of \(\An_\text{reg} ^{(D)} \) in \eqref{odj} was pointed out, corresponding to the dimensionality of the subtraction term, \(D=4\) vs.\ \(D=4-2\varepsilon\). We have reviewed and completed the calculation, spelling out some aspects that were misunderstood in the previous literature \cite{Casarin:2018odz,Ferrero:2023unz}. We explicitly showed that the prescription \(\An^{(4)}_\text{reg}\) reproduces the result of the heat-kernel identification \(\An_\text{hk} = a_4\), which is advocated in \cite{Duff:1993wm,Duff:2020dqb} to be preferred. On the other hand, the prescription \(\An^{(4-2\varepsilon)}\) is singled out in the analysis of \cite{Larue:2023qxw}. There, in the context of dimensional regularization, a different notion of \(\An\) valid to all-loop order is introduced, which is  by construction finite local and reduces to \(\An^{(4-2\varepsilon)}\) for free theories. It is naturally of interest to understand   which prescription is more appropriate to capture the sought effects.

In either case, \(\An \)  of the form \eqref{aaa} produces a quantity that contains \(R^2\), thus violating the WZ consistency conditions. This implies that it cannot be obtained as functional derivative of an effective action and it is not subject to the same counterterm ambiguity of the anomaly proper: a finite counterterm would cancel in the difference between the two terms in \(\An\). As an additional consequence, also the coefficient of \(\Box R\) is physical. Similar comments appear also in  \cite{Casarin:2021fgd,Larue:2023qxw}.

We have then extended the analysis of \cite{Casarin:2018odz,Casarin:2021fgd} to include interaction at lowest order in the coupling. We have shown that the regularized prescription is insufficient, as it gives a divergent result once the regulator is removed. We thus considered the definition \(\An_\text{ren}\) built of  renormalized correlators. We  have argued that it is a good candidate to extend the identification of \(\An\) with the heat kernel coefficient in the presence of interactions, at least for generic massless theories. This identification is nontrivial, in that it suggest a firmer diagrammatic understanding of the HK prescription  \eqref{aac} in a way that can be extended to higher  loops, and deserves to be investigated in greater generality.   

This definition, however, displays nonlocalities at higher loops. 
We have shown this explicitly in \eqref{fdo} and \eqref{fdr}.
We  explained the appearance of the nonlocalities as a consequence of the fact that, in constructing finite composite operators,  the stress-tensor  does not require any additional renormalization, while the operator associated to its trace does.  It is this imbalance that produces uncancelled nonlocal terms from two-loops on.

Given this discussion, it seems that the situation regarding the characterisation of quantum violation of Weyl invariance, when the classical symmetry is absent, is  far from clear. As Weyl (conformal) invariance is absent along the RG flow, this has the potential application of shedding light on the space of QFTs and providing insights in the local version of the \(a\) theorem. Similarly, Einstein gravity and supersymmetric generalisation thereof lack classical Weyl invariance, therefore the significance of the cancellation of the \(c\)-anomalies in the total heat kernel coefficients in \(N>4\) supergravities is unclear \cite{Meissner:2016onk,Meissner:2017qwm}.

On a practical level, it would be interesting to extend our calculation to higher loop to see the appearance of the beta functions as well as including mass terms. Other field theory  models  would provide additional concrete examples and would e.g.\ allow one to test the identification of \(\An\) with the heat kernel coefficient more thoroughly.    To make a clearer connection with the $a$-theorem \cite{Komargodski:2011vj}, it would  be interesting to compute  $\An_\text{ren}$ without   the spacetime integration  considered in section~\ref{sect:inta} that hides the contribution from $\EE$; more advanced diagrammatic techniques are needed in order to overcome the computational complexity. With this in mind, it would also be of interest understanding how to connect the notions of anomaly discussed above with  \cite{Jack:1990eb,Jack:2013sha,Baume:2014rla}.

\subsection*{Acknowledgments}
LC thanks H.\ Nicolai, S.\ Theisen and R.\ Zwicky for discussions and correspondence.

\appendix

\section{Notation and conventions}\label{app:not}  
We work in euclidean signature. Dimensional regularization is considered in  \(d=4-2\varepsilon\) dimensions. The metric is expanded in a perturbation around a flat background as \(g_{mn}(x)=\delta_{mn}+ h_{mn}(x)\).

Flat-space Fourier transforms and integrals follow the convention
\begin{equation}\label{nos}
f(x) = \int \dm{p} e^{ipx} f(p)
\;,
\qquad\quad
\int \dm{p} e^{ipx} = \delta(x)\,,
\qquad\quad
\dm{p} \! = \frac{d^d p }{(2\pi)^d}   \; .
\end{equation}
The four dimensional Euler density and Weyl curvature tensor  are given respectively by
\begin{equation}\label{fdc} 
\EE =   \Riem^2  - 4 \Ric^2 + R^2
\,,
\qquad\qquad
\WW =  \Riem^2 - 2\Ric^2 + \frac{1}{3} R^2
\,.
\end{equation}
Quantum expectation values are denoted as:

\(\braket{...}\): renormalized (finite) expectation values;

\(\braket{...}_{\mathcal O (h^n)}\): renormalized (finite) expectation values of order \(n\) in the metric perturbation;

\(\braket{...}_\text{reg}\):  regularized but not renormalized correlators (only used in general discussion);

\(\braket{...}_\varepsilon  \):  regularized   correlators  in dimensional regularization; 

\(\bb{... } _\varepsilon   \): bare correlators   taken in the   free theory, in flat space, in dimensional regularization.

\section{Feynman rules and diagram integrals}\label{app:vert}
 The propagator for the field \(\varphi\) in momentum space reads
\begin{equation}\label{pac}
 G(p,q) 
 =  \bb{ \varphi(p) \varphi(q) } _\varepsilon
 = \frac{  (2\pi)^d \ \delta^{(d)}[p+q]  }{p^2}  
\end{equation}
The action vertices as defined in \eqref{baa} and following, are:
\begin{equation}\label{kdc}
\begin{aligned}
V^{ (2)}_{mn,rs }(p)    & =
  \frac{\gamma}{ 2 }  p^4 \delta_{  m  ( r  }\delta_{  s  )  n  } 
-  \gamma   p^2 p_{ (m }\delta_{  n)  (  r  }   p_{ s )} 
+ p_m p_n  p_r p_s  \left(  \rho +   \frac{ \gamma } { 3 }  \right)
\\\nonumber
&\qquad{} 
+\left( p^2  p_r p_s \delta_{mn} 
+ p^2  p_m p_n \delta_{rs}\right) \left(   \frac {\gamma} 6    - \rho   \right)
- p^4 \delta_{  m  n  }\delta_{  r  s  } \left(   \frac  {\gamma} 6    - \rho \right)
\\[0.5em]
  V^{\varphi^2(1)}_{mn}(p,q,\ell)  & = 
  \frac 12 p_{ (m }  q_{  n )  } - \frac14 \delta_{mn} \,pq
  +\frac\xi2   \left(
  		\delta_{mn} \ell ^2 - \ell_{  m } \ell_{  n }
  \right)
\\[0.5em]
V^{\varphi^2(2)}_{mn,rs}(p,q,\ell,k)   & = 
	- \frac 1 {16} \delta_{mn} \delta_{rs} pq
	+ \frac 18 pq \delta_{ m (r  }  \delta_{ s) n    }
	- \frac14 p_{(m  } \delta_{  n ) (r  }  q_{  s) } 
	- \frac14 q_{(m  } \delta_{  n ) (r  }  p_{  s) } 
	+ \frac18   \delta_{ mn } q_{(r  }   p_{  s) } 
	+ \frac18   \delta_{ rs } q_{(m  }   p_{  n) } 
	\\* \nonumber
& \qquad {}
+ \xi \Big[
	\frac 18 \delta_{mn} \delta_{rs} (k^2+ k \ell+ \ell^2)
		- \frac 18 \delta_{  m (r  }  \delta_{ s ) n  } (  2 k^2 - 3 k \ell   + 2 \ell^2   )
			- \frac 18 \delta_{rs} (\ell_m \ell_n +2 k_m k_n )
\\* \nonumber
& \qquad\qquad {}
	- \frac 18 \delta_{mn} (k_r k_s 	+ 2 \ell_r \ell_s)
	- \frac 14 \delta_{mn} k_{ (r  }  \ell_{  s) }
	- \frac 14 \delta_{rs} k_{ (m  }  \ell_{  n) }
	+ \frac 12 k_{ (m  }   \delta_{  n )(r   }  k_{   s) }
	+ \frac 12 \ell_{ (m  }   \delta_{  n )(r   }  \ell_{   s) }
\\* 
& \qquad\qquad {}
	+ \frac 12  \ell_{ (m  }   \delta_{  n )(r   }  k_{   s) }
	+ \frac 14  k_{ (m  }   \delta_{  n )(r   }  \ell_{   s) }
\Big]
\\
V^{\varphi^4(1)}_{mn} &	= 	 \frac{\lambda}{2 \cdot 4!} \delta_{mn} 
\end{aligned}
\end{equation} 
We also use \(V^{ (3)}_{mnrsac }\), but its expression is lengthy and uninformative  so we  do not report it.

The stress tensor vertices as defined in \eqref{bcb} and following  are
\begin{equation}\label{das} 
\begin{aligned}
			   W^{\varphi^2(1)}_{mn}(p,q)  &= -2  V^{\varphi^2(1)}_{mn}(p,q,-p-q) \,,
\qquad  
	  W^{\varphi^4(0)}_{mn}  = -  \frac{\lambda}{4!}  \delta_{mn}\,,
\qquad  
 		  	  W^{\varphi^4(1)}_{mnac}   - \frac{\lambda}{4!}  \delta_{a(m}\delta_{n)c} \,,\!\!\!
\end{aligned}
\end{equation}

The integrals corresponding to the diagrams of figure \ref{fig:feyn1} referring to equation \eqref{fks} are 
\begin{align} 
\nonumber
- \bb{T_{mn}^{\varphi^2 (0)}      S^{ (1) }    _{ \varphi^2}   }_\varepsilon &=
-2 \int \dm{q} e^{iqx} h_{rs}(q) \int \dm{p} \frac{1}{p^2 (p-q)^2} 
W^{\varphi^2(0)}_{mn}(p,q-p)    \   V_{rs}^{\varphi (1)} (  -p,p-q,q  )
\\    \nonumber
\frac12 \bb{T_{mn}^{\varphi^2 (0)}     S^{ (1) }    _{ \varphi^2} S^{ (1) }    _{ \varphi^2}   }_\varepsilon
&=
4 \int \dm{k} \dm{\ell} e^{i(k+\ell )x} h_{ac}(\ell) h_{rs}(k)
\int \dm p \frac{1}{p^2 (p-\ell)^2 (p+k)^2}  \cdot {}
\\
&\qquad\qquad
{} \cdot 
W^{\varphi^2(0)}_{mn}(\ell-p,p+k) 
   \   V_{rs}^{\varphi (1)} (  p,-p-k,k  )
      \   V_{ac}^{\varphi (1)} (  p-\ell,-p,\ell  ) \nonumber \\
 - \bb{T_{mn}^{\varphi^2 (0)}     S^{ (2) }    _{ \varphi^2}   }_\varepsilon 
&=
- 2 \int \dm{k} \dm{\ell} e^{i(k+\ell )x} h_{ac}(\ell) h_{rs}(k)
		\int \dm{p} \frac{1}{p^2 (p-k-\ell)^2 }
		\\ \nonumber
  & \qquad \qquad \cdot W_{mn}^{\varphi^2(0)}  (p,k+\ell-p)
		 V_{acrs}^{\varphi^2(2)} (-p,p-k-\ell,\ell , k)  \nonumber \\ \nonumber
	- \bb{T_{mn}^{\varphi^2 (1)}     S^{ (1) }    _{ \varphi^2}   }_\varepsilon
&=
-2 \int \dm{k} \dm{\ell} e^{i(k+\ell )x} h_{ac}(\ell) h_{rs}(k)
		\int \dm{p} \frac{1}{p^2 (p-k)^2 } \\
  &\qquad \qquad \cdot W_{mnac}^{\varphi^2(1)}  (p,k-p,\ell)
		 V_{rs}^{\varphi^2(1)} (-p,p-k,k) \nonumber
\end{align}

The bare integrals corresponding to the diagrams of figure \ref{fig:feyn2} referring to equation \eqref{dcl} are 
\begin{align}\label{fsc}
T^{\text i \ \text{bare} }_{mnrsac}(k)
&=
- 2    
\int \dm{p } \frac{W^{\varphi^2 (1)}_{mnrs} (-p,-k+p,k)   V^{\varphi^2 (1)}_{ac} (p,k-p,-k)   }{p^2 (p-k)^2} 
\\ \nonumber
& -8
 \int \dm{p } \frac{W^{\varphi^2 (1)}_{mn} (-p,p,0)V^{\varphi^2 (1)}_{rs} (-p,p-k,k)   V^{\varphi^2 (1)}_{ac} (p,k-p,-k)  }{p^4 (p-k)^2} \\  \nonumber
&+ 12  \int \dm{p}\frac{ V_{rs}^{\varphi^2 (1)} (-p,p-k,k)}{p^2 (p-k)^2}  \int \dm{q} \frac{V_{ac}^{\varphi^2 (1)} (q,-q+k,k)}{ q^2 (q-k)^2}  W^{mn} (p,-q,-p+k,-k+q)
\\\nonumber
&   
+ \lambda  
\int \dm{p }  \frac{W^{\varphi^2 (1)}_{mnrs} (-p,p-k,k)  }{p^2 (p-k)^2} 
 \int \dm{q} \frac{V^{\varphi^2 (1)}_{ac} (q,-q+k,-k)  }{q^2 (q-k)^2} 
 \\\nonumber
  &- 2 \lambda  
 \int \dm{p } \frac{V^{\varphi^2 (1)}_{rs} (-p,p-k,k) }{ p^2 (p-k)^2  } 
  \int  \dm{q} \frac{ V^{\varphi^2 (1)}_{ac} (q,-q+k,-k) W_{mn}^{\varphi^2 (0)}(-q,q) }{   q^4  (q-k)^2    } 
\\[0.5em]
T^{\text {ii}  \, \text{bare}  }_{mnac}(k)
&=
4 \int \dm p \frac{ 
			V^{\varphi^2(1)}_{mn} (-p,p-k,k) 
				\,  V^{\varphi^2(1)}_{mn} (p,k-p,-k)
	 }{p^2 (p-k)^2} \label{fscc}
\\\nonumber
&
-2\lambda 
 \int \!  \dm{p }  
  \frac{V^{\varphi^2 (1)}_{mn} (-p,-k+p,k)  }{p^2 (p-k)^2}
  \int \! \dm q  \frac{ V^{\varphi^2 (1)}_{ac} (k-q,q,-k)  }{q^2 (q-k)^2}   \,,  
\end{align}
where we have only given the bare integrals corresponding to the diagrams displayed; the counterterm contributions can be easily derived.

\section{Remarks on  action renormalization on curved background}\label{app:renact}

In the notation explained in Section~\ref{sect:reg}, the bare theory induces purely gravitational infinities that need to be cancelled by counterterms in \(S_{\text{grav}}\) as in \eqref{bac} with \eqref{cfd} and \eqref{wsf}.
To determine the counterterms it is enough to compute the effective action to second and third order in the \(h\) expansion. 

For the two-point function  we have
\begin{equation}\label{fsx}
\begin{aligned}
\Gamma_{\mathcal O(h^2)}& = 
 \int \dm{q} h_{mn}(-q) h_{rs}(q)
 \Big\{ 
	\int \dm{p} \frac{1}{p^2(q-p)^2} 
				V^{\varphi^2(1)}_{mn}(p,q-p,-q)
			\	 V^{\varphi^2(1)}_{rs}(-p,p-q,q)
\\&\qquad {} - \frac{\lambda}{2}
	\int \dm{p} \frac{1}{p^2(q-p)^2} 
				V^{\varphi^2(1)}_{mn}(p,q-p,-q)
	\int \dm{k} \frac{1}{k^2(k-p)^2} 
			\	 V^{\varphi^2(1)}_{rs}(-k,k-q,q)
			 \Big\}.\!\!\!
\end{aligned}
\end{equation} 
Performing the calculation to two loop (first order in \(\lambda\)) fixes \(\xi^{(1)}, \xi^{(2)}\) through subdiagrams, and the resulting divergences give the counterterms \(\gamma^{(1)}\),    \(\gamma^{(2)}\),  \(\rho^{(1)}\) and \(\rho^{(2)}\). In contrast, \(\alpha\) is not captured because \(\EE\) does not have a quadratic term in the expansion on a flat background in general dimension.

The bare three point  function is 
\begin{equation}\label{fsy}
\begin{aligned}
\Gamma_{\mathcal{O}(h^3)} &= 
\int \dm{p}\dm{q} h_{mn}(-p) h_{rs}(q) h_{ac}(p-q)
\Big\{
 \frac43 
	\int \dm{\ell} \frac{1}{\ell^2 (\ell-p)^2(\ell-q)^2}  \times
\\
& \qquad\qquad\qquad\qquad
				\times
				V^{\varphi^2(1)}_{mn}(\ell,p-\ell,p) 
				V^{\varphi^2(1)}_{rs}(-\ell,\ell-q,q) 
				V^{\varphi^2(1)}_{ac}(\ell-p,q-\ell,p-q) 
\\
& \qquad
{}-2 
	\int \dm{\ell} \frac{1}{\ell^2(\ell-p)^2} 
				V^{\varphi^2(1)}_{mn}(\ell,p-\ell,-p)
			\	 V^{\varphi^2(2)}_{rsac}(-\ell,\ell-p,q,p-q) 
	\Big\}.
\end{aligned}
\end{equation}
As observed in \cite{Goroff:1985th}, the three point function does capture the coefficient of  $\EE$. Despite the fact that it is a total derivative in \(D=4\) and vanishes in \(D<4\), in the spirit of analytically continuing for generic (complex) \(D\), it is indeed relevant in the \(\varepsilon\) expansion. In fact, the contribution disappears only by using identities  that are not valid for \(D>4\).

\bibliography{biblio} 

\bibliographystyle{JHEP}

 \end{document}